\documentclass[prl,twocolumn,showpacs,amsmath,amssymb,superscriptaddress]{revtex4}

\usepackage{graphicx}
\usepackage{dcolumn}
\usepackage{url}
\usepackage{amsmath}
\usepackage{bm}
\usepackage{latexsym}

\begin{document}

\title{Efficient Quantum State Estimation by Continuous Weak Measurement and Dynamical Control}

\author{Greg A. Smith}
\affiliation{College of Optical Sciences, University of Arizona, Tucson, AZ 85721}
\author{Andrew Silberfarb}
\affiliation{Department of Physics and Astronomy, University of New Mexico,Albuquerque, NM 87131}
\author{Ivan H. Deutsch}
\affiliation{Department of Physics and Astronomy, University of New Mexico,Albuquerque, NM 87131}
\author{Poul S. Jessen}
\affiliation{College of Optical Sciences, University of Arizona, Tucson, AZ 85721}

\date{\today}

\begin{abstract}
We demonstrate a fast, robust and non-destructive protocol for
quantum state estimation based on continuous weak measurement in the
presence of a controlled dynamical evolution.  Our experiment uses
optically probed atomic spins as a testbed, and successfully
reconstructs a range of trial states with fidelities of $\sim90\%$.
The procedure holds promise as a practical diagnostic tool for the
study of complex quantum dynamics, the testing of quantum hardware,
and as a starting point for new types of quantum feedback control.
\end{abstract}

\maketitle



Fast, accurate and robust quantum state estimation (tomography) is
important for the study of complex quantum systems and dynamics, and
promises to be an essential tool in the design and testing of
hardware for quantum information processing \cite{NielsenChuang}.
Previous demonstrations range from optical \cite{PRLSmithey} to
atomic \cite{PRLLeibfried} and molecular systems \cite{PRLWalmsley},
but with few exceptions these procedures have proven too cumbersome
to be of use as practical laboratory tools. The procedure of quantum
state estimation is usually formulated in terms of strong
measurements of an informationally complete set of observables. Each
such measurement erases the original quantum state, so the ensemble
must be reprepared and the measurement apparatus reconfigured at
each step.  Here we demonstrate a general approach based instead on
continuous weak measurement \cite{PRLSilberfarb}. Using a weak
measurement spreads quantum backaction across the ensemble
and dilutes it to the point where it does not significantly affect
any individual member.  In the absence of backaction the quantum state
remains largely intact, subject only to minimal damage from errors
in the external drive fields and coupling to the environment.  This
allows us to estimate the state in a single interrogation of the
ensemble, based on the measurement of a fixed observable $O$ and a
carefully designed system evolution.

In the Heisenberg picture the chosen dynamics leads to a
time-dependent observable, $O\rightarrow O(t)$, and the measurement
history can be made informationally complete if the system is
\emph{controllable}, i.\ e.\ the dynamics can generate any unitary
in $SU(d)$ where $d$ is the dimensionality of Hilbert space. With a
non-destructive measurement and near-reversible dynamics, the entire
ensemble remains available at the end of the estimation procedure,
in a known quantum state that can be restored close to its initial
form if desired.  In principle, this means that the knowledge gained
can be used as a basis for further action, for example real-time
feedback control \cite{PRADoherty} or error correction
\cite{NielsenChuang}. From a practical viewpoint our procedure is
highly efficient: the $\sim1ms$ interrogation time is limited only
by the control and measurement bandwidths, and data analysis is
performed offline.  It is also robust, in the sense that
imperfections in the experiment can be included in the analysis if
known, or estimated along with the state if they fluctuate in real
time.

\begin{figure*}[tbp]
\includegraphics[width=7in]{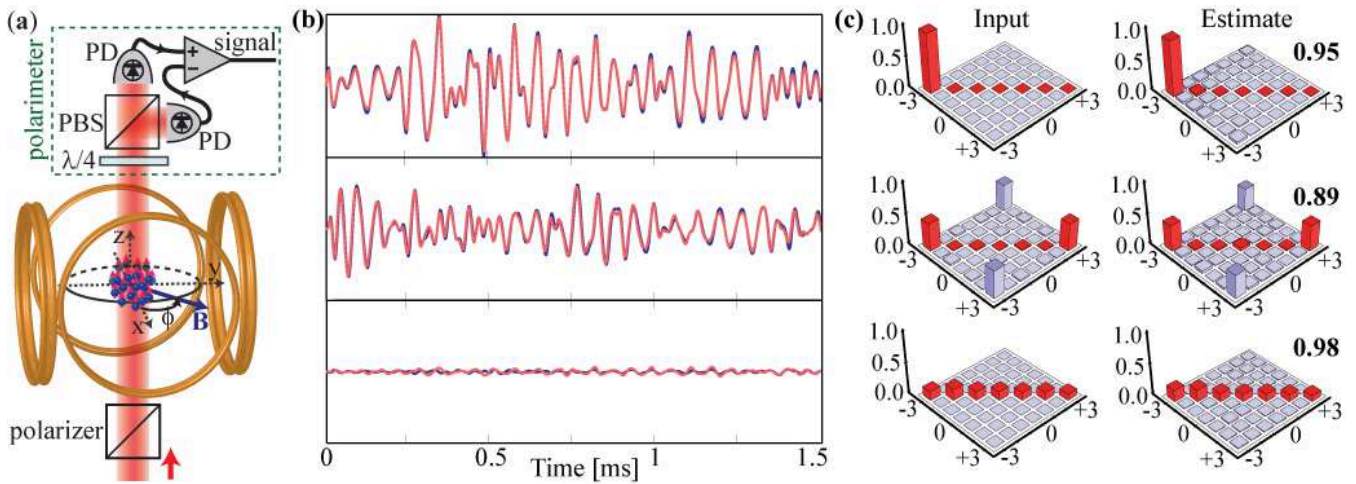}
\caption{\label{fig-expt}Continuous measurement and quantum state estimation for
atomic spins. (\textbf{a}) Schematic of our experiment. PBS:
polarization beamsplitter; PD: photodetector; $\lambda\, /\, 4$:
quarter-wave plate. The control magnetic field has constant
magnitude and is confined to the x--y plane; the time dependent
angle $\phi$ is optimized for best estimation fidelity.
(\textbf{b})Simulated (dark blue) and observed (light red)
measurement signals for test states $\vert m_{F}=-3\rangle$ (top),
$\vert\psi_{c}\rangle$(middle) and $\rho_{mix}$ (bottom).
(\textbf{c}) Input and estimated density matrices (absolute values)
corresponding to the simulated and observed measurement signals.
(Color online)}
\end{figure*}

The quantum system used in our laboratory implementation is the
total spin-angular momentum (electron plus nuclear) of an atom in
the electronic ground state, here the $F=3$ hyperfine manifold of
$^{133}$Cs. A detailed theoretical description of our estimation
procedure and how to model this system can be found in
\cite{PRLSilberfarb}. Our Cs ensemble is prepared by laser cooling
and optical pumping, and consists of a few million atoms in a cloud
with radius $\sim0.3\,mm$. The spins are coupled to the polarization
of an off-resonance probe beam that serves as a meter for the
measurement and can be read out in an arbitrary basis with a
shot-noise limited polarimeter \cite{JOptBSmith}, as illustrated in
fig.\ \ref{fig-expt}a.
The spin-probe coupling depends on the atomic tensor polarizability,
and can be expressed in terms of the single-atom spin  \textbf{F}
and the probe Stokes vector \textbf{S},
\begin{align}
V_{SP} &= \xi _{2}\,[ -F_{z}^{2}\,S_{0} + \left(
F_{x}^{2}-F_{y}^{2}\right)\,S_{1}\nonumber\\
&+\,\left( F_{x}\,F_{y}
+F_{y}\,F_{x}\right) \,S_{2}]+\xi _{1}\,F_{z}\,S_{3}
\label{eqn-Stokes}
\end{align}

where the coupling constants $\xi _{i}$ depend on the probe
frequency, the cloud optical density, and the atomic transition
\cite{unpubDeutsch}. Terms containing $S_{1}$, $S_{2}$, and $S_{3}$
lead to rotation of the Stokes vector around the 1, 2 and 3-axis of
the Poincar\'{e} sphere, by angles proportional to the
ensemble-averaged values of the respective spin observables. We use
a probe initially polarized along the $x$-axis,
corresponding to an input Stokes vector $\textbf{S}_{in}=S_{0}\,\left( 1,0,0\right) $%
, and measure the induced ellipticity, or $S_{3}$\ component of the
output. For the moderate optical density clouds used here all angles
of rotation are small, the polarization measurement is sensitive
only to rotations around the 2-axis, and the measured observable is
$O=F_{x}\,F_{y}+F_{y}\,F_{x}$. Coupling the entire spin ensemble to
a common probe can, in principle, produce correlations between
individual spins \cite{SCIGeremia}, but this effect is negligible
for our optical densities.

To extract information we evolve the spins to obtain a time varying
observable $O\left( t\right) $. Coarse-graining over time yields a
discrete set $\left\{ O_{i}\right\} $\ and a measurement record $
\left\{ M_{i}\right\} $, where $M_{i}=$Tr$\left[ O_{i}\,\rho
_{0}\right] +\Delta M$. Here $\rho _{0}$ is the quantum state to be
estimated, and $\Delta M$ is a Gaussian white noise process. Full
system controllability is required to make $\left\{ O_{i}\right\} $
informationally complete, and can be achieved with a Hamiltonian
of the form $H_{s}\left( t\right) =g_{f}\,\mu _{B}\,\mathbf{B}\left(
t\right) \cdot \mathbf{F}+\beta \,\hbar \,\gamma _{s}\,F_{x}^{2}$
produced by a time-dependent magnetic field and the probe-induced
tensor light shift (eq. \ref{eqn-Stokes}) \cite{PRLSmith}. Note that
for alkali atoms with an $nS_{1/2}$ electronic ground state, the
tensor light shift is proportional to the photon scattering rate
$\gamma _{s}$ and therefore comes at the cost of decoherence. By
probing Cs atoms on the D$_{1}$ line we achieve $\beta =8.2$, enough
to allow significant coherent evolution before optical pumping
affects the spin state. We model our system dynamics with a master
equation that fully accounts for optical pumping and experimental
imperfections. The latter include bandwidth limits for our magnetic
coil drivers and detection system, a 6\% spatial inhomogeneity
in the magnitude of the light shift which can be independently
estimated, and a $\sim1\%$ drift in our magnetic coil calibrations
which must be included as free parameters in the estimation
procedure.

We have evaluated the performance of our state estimation procedure
by applying it to a range of test states. Nearly pure test states
with $m_{F}=0,\,\pm F$ were produced by optical pumping, and
verified by Stern-Gerlach measurements of the magnetic populations.
Geometric rotations of these states were obtained by Larmor
precession, and more general test states were produced by evolving
the system with a magnetic field and the tensor light shift. A
strongly mixed state was obtained by omitting the optical pumping
step and working directly with the laser cooled sample.
Figure \ref{fig-expt}b compares the measurement record observed in a
single ensemble interrogation to the one predicted by our model, for
three representative inputs corresponding to a spin-polarized state
$\left\vert m_{F}=-3\right\rangle $, a coherent superposition
$\left\vert \psi _{c}\right\rangle =\frac{1}{\sqrt{2}}\,\left(
\left\vert m_{F}=3\right\rangle +i\,\left\vert m_{F}=-3\right\rangle
\right) $, and a nearly maximally mixed state $\rho _{mix}\sim
I\,/\,7$.  Two key features are immediately apparent: the measurement records
differ substantially for different input states, and the observed
and calculated measurement records agree very closely. Both
conditions are clearly necessary for the procedure to generate
unique and accurate state estimates.

Based on our model we can determine the probability distribution for
the measurement record, $P\left( \left\{ M_{i}\right\} \,|\,\left\{
\rho _{0},\,p _{k}\right\} \right) $, conditioned on the initial
quantum state $\rho _{0}$ and any unknown experimental parameters
$p_{k}$. We use this to obtain a maximum likelihood estimate $\rho
_{ML}$ for the state, corresponding to the peak of the posterior
distribution, $P( \left\{ \rho _{0},\,p _{k}\right\}\,|\,\left\{
M_{i}\right\} ) = A\,P\left( \left\{ M_{i}\right\} \,|\,\left\{ \rho
_{0},\,p _{k}\right\} \right) \,P\left( \left\{ \rho
_{0},\,p_{k}\right\} \right)$, where $P\left( \left\{ \rho
_{0},\,p_{k}\right\} \right) $ is the prior information and $A$ is a
normalization constant. Here we include as prior information only
that $\rho _{0}$ must be a physical state, i.\ e.\ a positive
Hermitian operator with unit trace. The conditional distribution
$P\left( \left\{ M_{i}\right\} \,|\,\left\{ \rho _{0},\,p
_{k}\right\} \right) $ is a multi-parameter Gaussian centered on
$\rho _{LS}$, the ordinary least-squares fit to the measurement
record. Its covariance matrix determines the uncertainty in each of
the $\left( 2\,F+1\right) ^{2}$ real-valued parameters required to
specify $\rho _{0}$, and consequently how well an arbitrary state
can be estimated. In setting up the procedure, the control magnetic
field was optimized to minimize these widths and reduce their
sensitivity to errors. In
practice the state estimate is obtained from the measurement record in a two-step process. First we determine the least-squares fit $%
\rho _{LS}$, which is frequently non-positive due to mismatch
between the modeled and actual dynamics, and then we find the
positive density matrix that lies closest to $\rho _{LS}$. This is
our maximum likelihood estimate $\rho _{ML}$. Figure \ref{fig-expt}c
compares density matrices for the input and maximum likelihood
estimates for each of our three representative test states. Visual
inspection clearly shows the excellent agreement achieved by our
procedure. A more quantitative performance measure can be obtained
by calculating the fidelity $\mathcal{F} =\left( \text{Tr}\sqrt{\rho
_{0}^{1\,/\,2}\,\rho _{ML}\,\rho _{0}^{1\,/\,2}}\right) ^{2}$
\cite{RepMathPhys} which for these three states fall in the range
0.89 -- 0.98. Averaging over our entire sample of test states we
achieve a mean fidelity $\overline{\mathcal{F}} \approx 0.86$\ for
estimates based on a single measurement record, and a slightly
improved $\overline{\mathcal{F}} \approx 0.91$\ for estimates based
on the average of 128 measurement records.

\begin{figure}[tbp]
\centering
\includegraphics[width=3.4in]{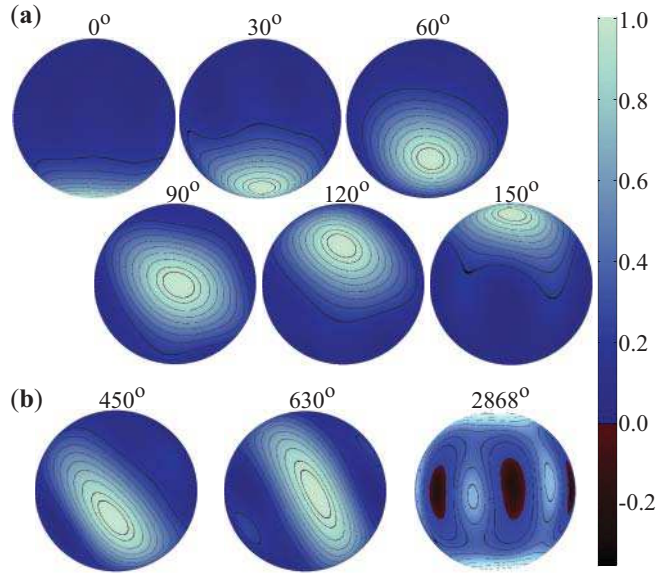}
\caption{\label{fig-wigner}Wigner function representations of an evolving spin state.
(\textbf{a}) Snapshots of the quantum state estimate as an initial
$\left\vert m_{F}=-3\right\rangle $ state undergoes Larmor
precession in the presence of the probe. The nominal precession
angle $\theta$ is indicated in each case. (\textbf{b}) Pronounced
squeezing of the spin wavepacket occurs at later times, and at
$\theta = 2868^{\circ}$ the system has evolved into a coherent
superposition $|\psi_{c}\rangle$ of spin-up/down wavepackets. The
viewpoint is fixed, except for the snapshots at $\theta =
630^{\circ}$ where it has been changed to show the wavepacket on the
opposite side of the sphere, and at $\theta = 2868^{\circ}$ where it
has been adjusted to show both lobes of $|\psi_{c}\rangle$.
(Color online)}
\end{figure}

An alternative visualization of spin states can be given in terms of
Wigner function representations \cite{PRAAgarwal}. Figure
\ref{fig-wigner}a shows the estimated states for six discrete time
steps during Larmor precession of an initial state $\left\vert
m_{F}=3\right\rangle $.
As expected for these so-called spin-coherent states, the Wigner
functions resemble Gaussian ``wavepackets'' moving on a sphere. A
slight but visible sheering of the quasi-probability distribution
occurs due to the $F_{x}^{2}$ term in the probe-induced light shift,
which is expected since this is a well known mechanism for spin squeezing \cite%
{PRAKitagawa}. Figure \ref{fig-wigner}b shows estimated states at
later times when the effects of squeezing are more pronounced. As
the squeezing ellipse wraps around the sphere it eventually evolves
into the coherent superposition $\left\vert \psi _{c}\right\rangle $
\cite{Sanders}, the state used as input in fig. \ref{fig-expt}b\&c. This series of snapshots demonstrates
that our technique can be used to observe and visualize quite
complex dynamical evolution.

\begin{figure}[tbp]
\centering
\includegraphics[width=3.4in]{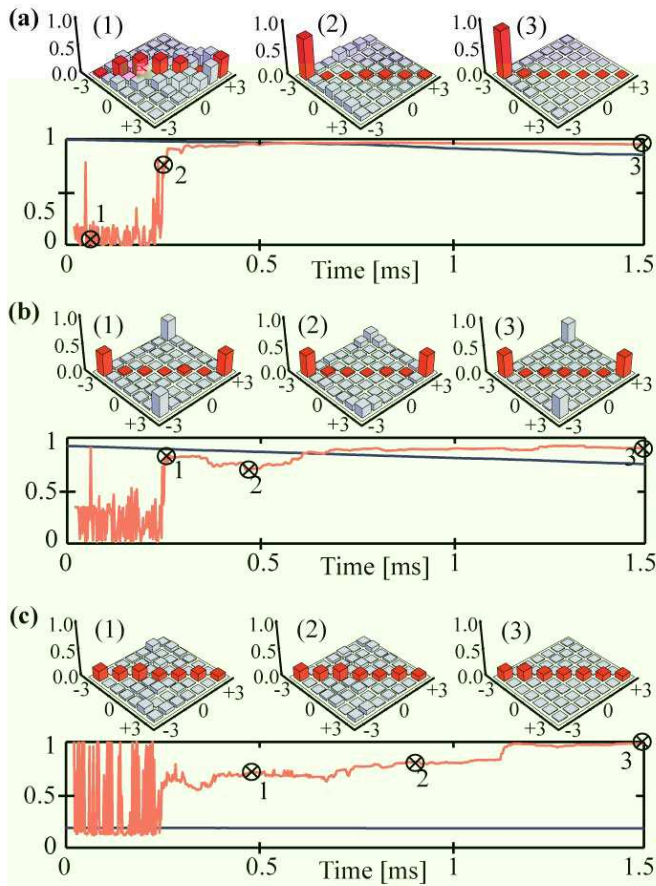}
\caption{ \label{fig-fidplot}Evolving fidelity, state estimate and purity. Plots show
the fidelity (light red) of the estimate $\rho_{ML}$ and the largest
eigenvalue (dark blue) of the evolving state $\rho_{0}$ as a
function of elapsed measurement time, for inputs (\textbf{a})
$\left\vert m_{F}=-3\right\rangle $, (\textbf{b}) $\left\vert
\psi_{c}\right\rangle $ and (\textbf{c}) $\rho _{mix}$. Inserts show
the estimated density matrices (absolute values only) at a few
representative times. (Color online)}
\end{figure}

With known input states we can observe how the state estimate
converges as the measurement record builds up. Figure
\ref{fig-fidplot} shows examples of the evolving fidelity and state
estimate for pure and mixed inputs.
In all cases, the estimation starts with an unbiased guess,
corresponding to a maximally mixed state. At short times, the
dynamics has explored only a small part of the set $\left\{
O_{i}\right\} $, and the measurement record provides good
information only about a few of the parameters that go into
specifying the state. As a result the estimate fluctuates randomly
with each time step, subject to noise and model/experiment
discrepancies. With increasing time, the dynamics explores a larger
set of observables, and the probability distributions are narrowed
for every parameter. At approximately $250\,\mu s$ the constraints
become strong enough that only a limited family of similar density
matrices are consistent with the measurement record, at which point
the estimate $\rho_{ML}$ stabilizes and the fidelity improves.
Figures \ref{fig-fidplot}a\&b show the behaviors for inputs
$\left\vert m_{F}=-3\right\rangle $ and $\left\vert \psi
_{c}\right\rangle $, where the transition at $250\,\mu s$ brings the
fidelity close to its final value. The details of the jump depend on
the semidefinite program solver used to enforce positivity, but it
is always dramatic for pure states where the positivity requirement
excludes most nearby points in parameter space. Fig.
\ref{fig-fidplot}c shows the behavior for an input $\rho_{mix}$,
where most nearby points in parameter space fulfill the positivity
requirement. As a result the initial jump is less dramatic and the
fidelity increases more gradually to its maximum value at the end of
the measurement period.

In the Heisenberg picture the state $\rho _{0}$ is time independent,
and the estimate $\rho _{ML}$ evolves only because more information
becomes available
in the measurement record. Once the measurement is complete the estimate $%
\rho _{ML}$ also becomes time independent. Because the dynamics are
known we can then transform to the Schr\"{o}dinger picture, where
the estimate becomes valid for any time, $\rho _{0}\left( t\right)
\approx \rho _{ML}\left( t\right) $. During the measurement period
there is, of course, a gradual and irreversible --- but entirely
predictable --- loss of information from the randomizing effects of
optical pumping and drive field errors. As a result an initially
pure state will become mixed, and the damage can be quantified by
keeping track of the largest eigenvalue of $\rho _{0}\left( t\right)
$. Figure \ref{fig-fidplot} shows a small decline for the input
$\sim\left\vert m_{F}=-3\right\rangle $, a slightly larger decline
for the more fragile superposition $\left\vert \psi
_{c}\right\rangle $, and essentially no damage to the mixed state
which is already randomized. Irrespective of the input state,
optical pumping transfers $\sim10\%$ of the ensemble into the $F =
4$ manifold during the $1.5\,ms$ interrogation, but these atoms are
invisible to the probe and do not contribute to the estimate or its
purity.

With this work we have shown that fast, accurate and non-destructive
quantum state estimation can be achieved with continuous weak
measurement and dynamical control. In our current implementation the
estimation fidelity is limited by imperfect control of the slowly
varying magnetic field and tensor light shift used to drive the
atomic spins. Our experience suggests that it may be possible to
achieve far more precise control of the atomic ground state by
driving the ensemble with radio-frequency and microwave radiation in
the presence of a static bias magnetic field. This would eliminate
optical pumping as a source of decoherence and loss, further reduce
the modest loss of state purity, and make the evolution almost
completely reversible. It could also allow robust state estimation
without invoking extraneous parameters and the positivity
requirement for $\rho_{ML}$, and this could shorten our data
processing time from currently $\sim1\,/\,2$ hour to perhaps a few
seconds on a desktop computer.  At that point, further improvements
in the algorithm and computing hardware --- for example the use of
programmable logic devices \cite{JOSABStockton} --- could plausibly
allow quantum state estimation and state-based feedback control in
real time.
\begin{acknowledgments}
This work was supported by the NSF (PHY-0355073 and PHY-0355040), ONR
(N00014-05-1-0420) and NSA/DTO.
\end{acknowledgments}

\end{document}